\documentclass[%
 reprint,
 amsmath,
 amssymb,
 aps,
 twocolumn,
 superscriptaddress
]{revtex4-1}

\usepackage{upgreek}

\makeatletter
\newcommand*{\rom}[1]{\expandafter\@slowromancap\romannumeral #1@}
\makeatother
\usepackage{amsmath}
\usepackage{graphicx}
\usepackage{bm}
\usepackage[caption=false]{subfig}
\usepackage{float}
\usepackage{placeins}
\usepackage{xcolor}
\usepackage{siunitx}
\usepackage{titlesec}
\usepackage{gensymb}
\usepackage{xcolor}
\usepackage{braket}

\titleformat{\section}
    {\normalfont\itshape\filcenter}{\thesection}{1em}{}
\titlespacing*{\section}{0pt}{0.5\baselineskip}{\baselineskip}

\begin{document}

\preprint{APS/123-QED}

\title{Hybridization gap approaching the two-dimensional limit of topological insulator Bi$_x$Sb$_{1-x}$}

\author{Paul Corbae}
\affiliation{Department of Electrical and Computer Engineering, University of California, Santa Barbara, California, 93106, USA}
\affiliation{Quantum Foundry, University of California, Santa Barbara, California, 93106, USA}
\thanks{To whom correspondence should be addressed; Email:pjcorbae@ucsb.edu; cjpalm@ucsb.edu}
\author{Aaron N. Engel}
\affiliation{Department of Materials, University of California, Santa Barbara, California, 93106, USA}
\affiliation{Quantum Foundry, University of California, Santa Barbara, California, 93106, USA}
\author{Jason T. Dong}
\affiliation{Department of Materials, University of California, Santa Barbara, California, 93106, USA}
\affiliation{Quantum Foundry, University of California, Santa Barbara, California, 93106, USA}
\author{Wilson J. Y\'{a}nez-Parre\~{n}o}
\affiliation{Department of Electrical and Computer Engineering, University of California, Santa Barbara, California, 93106, USA}
\author{Donghui Lu}
\affiliation{Stanford Synchrotron Radiation Lightsource, SLAC National Accelerator Laboratory, 2575 Sand Hill Road, Menlo Park, California 94025, USA}
\author{Makoto Hashimoto}
\affiliation{Stanford Synchrotron Radiation Lightsource, SLAC National Accelerator Laboratory, 2575 Sand Hill Road, Menlo Park, California 94025, USA}
\author{Alexei Fedorov}
\affiliation{Advanced Light Source, Lawrence Berkeley National Lab, Berkeley, California, 94702, USA}
\author{Chris J. Palmstr$\o$m}
\affiliation{Department of Electrical and Computer Engineering, University of California, Santa Barbara, California, 93106, USA}
\affiliation{Quantum Foundry, University of California, Santa Barbara, California, 93106, USA}
\affiliation{Department of Materials, University of California, Santa Barbara, California, 93106, USA}

\date{\today}

\begin{abstract}
Bismuth antimony alloys (Bi$_x$Sb$_{1-x}$) provide a tuneable materials platform to study topological transport and spin-polarized surface states resulting from the nontrivial bulk electronic structure. In the two-dimensional limit, it is a suitable system to study the quantum spin Hall effect. 
In this work we grow epitaxial, single orientation thin films of Bi$_x$Sb$_{1-x}$ on an InSb(111)B substrate down to two bilayers where hybridization effects should gap out the topological surface states. Supported by a tight-binding model, spin- and angle-resolved photoemission spectroscopy data shows pockets at the Fermi level from the topological surface states disappear as the bulk gap increases from confinement. Evidence for a gap opening in the topological surface states is shown in the ultrathin limit. Finally, we observe spin-polarization approaching unity from the topological surface states in \SI{10}{} bilayer films. The growth and characterization of ultrathin Bi$_x$Sb$_{1-x}$ alloys suggest ultrathin films of this material system can be used to study two-dimensional topological physics as well as applications such as topological devices, low power electronics, and spintronics.

\end{abstract}

\maketitle

\section*{Introduction}
Bulk Bismuth antimony alloys (Bi$_x$Sb$_{1-x}$) provided the first experimental platform to observe the surface states of a strong three-dimensional topological insulator \cite{doi:10.1126/science.1167733,Hsieh2008,PhysRevB.78.045426}. Bi and Sb are group five semimetals with a direct gap throughout the entire Brillouin zone but a negative indirect gap from band overlap. At the $L$ point in Bi the valence band (VB) maximum consists of odd parity antisymmetric orbitals and the conduction band (CB) minimum consists of even parity symmetric orbitals with $Z_2=(0;000)$. In pure Sb the orbitals are inverted leading to a nontrivial bulk invariant $Z_2=(1;111)$. By alloying Sb into Bi, the small gap at the $L$ point inverts and opens leading to a topologically nontrivial bulk insulator \cite{PhysRevB.78.045426}. The physical responses of topologically insulating systems makes them suitable materials platforms for low-power electronics, spintronics, and quantum information because the topological surface states (TSS) have their spin locked perpendicular to momentum and the nontrivial Berry phase prevents backscattering\cite{RevModPhys.82.3045,Tokura2017}. When the dimensionality is reduced the top and bottom surface states of a 3D TI can hybridize and the Kramers degeneracy is removed\cite{Zhang2010}. Due to the gap in the surface states and nontrivial bulk topological invariant, the potential for another quantum phase of matter emerges, the quantum spin Hall effect (QSHE) \cite{doi:10.1126/science.1133734}. In bulk form the small electronic band gap of \SIrange{10}{30}{meV}\cite{PhysRevMaterials.5.L091201,PhysRevB.52.1566} makes isolating the topological surface states of Bi$_x$Sb$_{1-x}$ from parasitic effects in transport difficult. Addressing this problem is important for the use of Bi$_x$Sb$_{1-x}$ in topological devices, which motivates reducing the dimensionality of this material system by growing in thin film form. In thin films, the electrons are confined and the spacing of the discrete energy levels increases, leading a larger effective bulk energy gap \cite{Zhang2010}. Understanding the electronic structure and spin-polarized surface states in Bi has intrigued many, however ultrathin Bi grows in different phases depending on thickness and substrate, stabilization of (111) oriented films below a few nanometers is difficult \cite{Schindler2018,Drozdov2014,HOFMANN2006191,Takayama2012,doi:10.1126/science.aai8142,PhysRevLett.93.105501,inbar2023inversion,Chou2022,PhysRevMaterials.6.074204,ou2023spin,huang2023epitaxial,Khang2018,doi:10.1126/sciadv.aay2324,https://doi.org/10.1002/advs.202301124,PhysRevMaterials.6.074204,10.1063/5.0190217}. In addition to the $Z_2 = 1$ phase of Sb \cite{doi:10.1126/science.1167733} and higher order topology of Bi \cite{Schindler2018}, the stabilization of a few bilayer film enables Bi$_x$Sb$_{1-x}$ thin films to be used in studying the emergent physics of two dimensional systems such as the QSHE, quantum anomalous Hall effect, two-dimensional Dirac semimetals, or 2D Weyl fermions \cite{doi:10.1126/science.aai8142,PhysRevB.91.041303,Kowalczyk2020,lu2023observation}.

\begin{figure*}
\centering
  \includegraphics[width=1\textwidth]{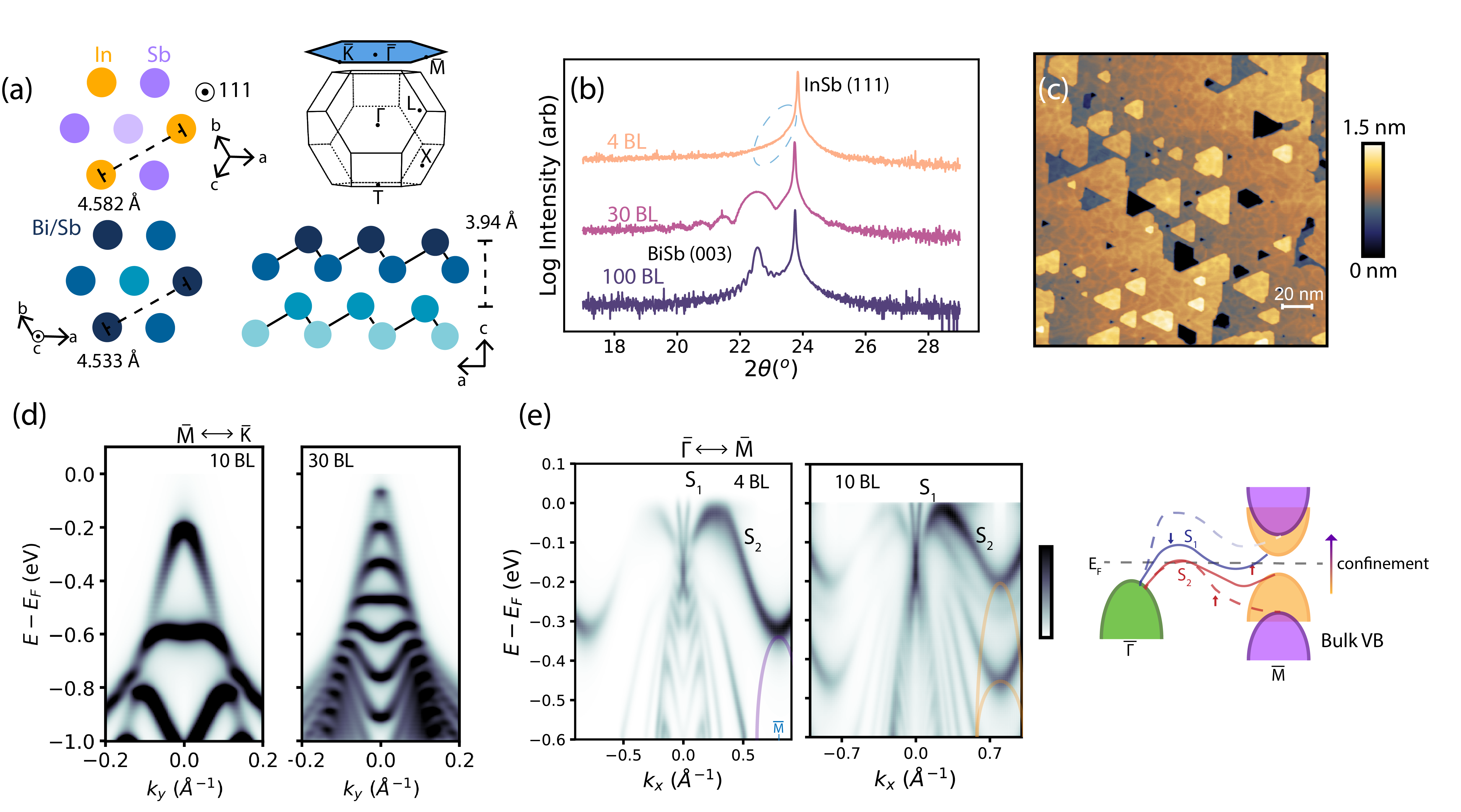}
  \caption{\textbf{Real and momentum space structure of Bi$_{0.85}$Sb$_{0.15}$ grown on InSb(111)B.} \textbf{(a)} The crystal structure of InSb and Bi$_{1-x}$Sb$_{x}$. Top view of the InSb(111)B and Bi$_{1-x}$Sb$_{x}$(0001) surface. The BiSb alloy is well lattice matched to the InSb substrate. The side view of the BiSb alloy shows the Van der Waals character of the system. Schematic of the bulk Brillouin zone. The projection onto the (111) surface shows the bulk $L$ point projects onto the surface $M$ point. \textbf{(b)} XRD spectra for \SI{4}{}, \SI{30}{}, and \SI{100}{BL}. The BiSb (0003) peak is parallel to the InSb(111) peak and the spacing of the finite size lattice fringes decreases with increasing thickness. The fringe spacing for each curve reflects the film thickness. There is a shoulder as $2\theta$ decreases from the InSb(111) peak from the \SI{4}{BL} film (circled). \textbf{(c)} STM image of \SI{4}{BL} Bi$_{0.85}$Sb$_{0.15}$ at bias voltage of \SI{500}{mV} and current \SI{500}{pA}. Single orientation terraces can be seen as well as strain solitons. \textbf{(d)} Simulated ARPES intensity for \SI{10}{} and \SI{30}{BL} slabs in the $\Bar{K}-\Bar{M}-\Bar{K}$ direction with $p$-polarized light at \SI{14}{K}. As the film thickness is increased, the bulk VB at the L point moves closer to the Fermi level as the bulk gap decreases. \textbf{(e)} As the film thickness is increased $S_2$ connects to the VB at smaller binding energies and the QW spacing decreases as the film thickness is increased. Bulk VB QWs sketched on top of spectral function. The effect of confinement on the TSS and electronic structure of Bi$_{0.85}$Sb$_{0.15}$ is summarized by the cartoon.}
  \label{fig:structure}
\end{figure*}

In this work we grow epitaxial thin films of Bi$_x$Sb$_{1-x}$ on a well lattice matched III-V semiconductor InSb(111)B substrate via molecular beam epitaxy (MBE). The structure of the films is investigated via X-ray diffraction (XRD) and scanning tunneling microscopy (STM). The films are seen to grow epitaxially, in a single orientation. The electronic structure is studied using angle- and spin- resolved photoemission spectroscopy (ARPES/SARPES) which is supported by ARPES simulations using a Slater-Koster tight binding model incorporating the matrix element effect. By reducing thickness we observe quantum well (QW) states and the separation of the TSS increases in energy from the confinement effect, inline with increasing the bulk gap. 
Line cuts from high resolution ARPES measurements show the width of the Dirac node increase in energy, inline with the opening of a hybridization gap. Finally, SARPES is used to show the topological states spin-polarization approaches unity then decreases at reduced thickness from hybridization. 
Our results highlight Bi$_x$Sb$_{1-x}$ in the ultrathin limit as a suitable system to functionalize the TSS for use in devices and for use in studying topological phases in two dimensions. By choosing a suitable substrate, we believe that we have enabled the thinnest, high-quality growth of Bi$_x$Sb$_{1-x}$ reported and study the TSS from bulk to few bilayer form with MBE.

\section*{Results}

\begin{figure*}
\centering
  \includegraphics[width=1\textwidth]{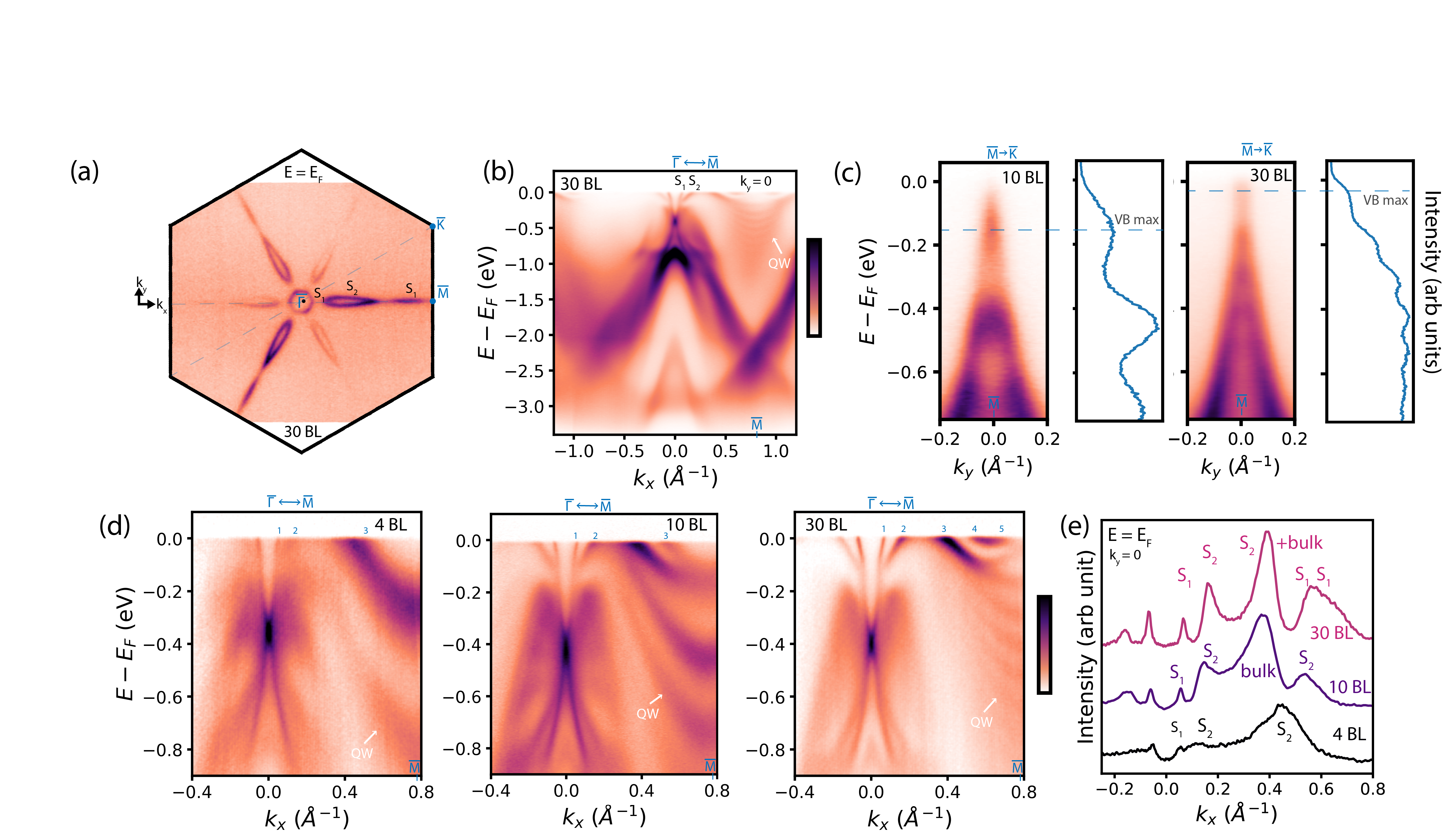}
  \caption{\textbf{Thickness dependence of the electronic structure in  Bi$_{0.85}$Sb$_{0.15}$ thin films.} All data at \SI{6}{K}. \textbf{(a)} Fermi surface for \SI{30}{BL} Bi$_{0.85}$Sb$_{0.15}$ taken at $h\nu = \SI{37.5}{eV}$. The Fermi surface is three-fold rotationally symmetric from the bulk symmetry. $S_1$ can be seen to cross the Fermi level three times while $S_2$ crosses twice. \textbf{(b)} Wide energy $\Bar{\Gamma}-\Bar{M}$ cut for \SI{30}{BL} Bi$_{0.85}$Sb$_{0.15}$ at $h\nu = \SI{37.5}{eV}$. Both $S_1$ and $S_2$ can be seen in the \SI{30}{BL} film crossing five times total. There are obvious QW states $\Bar{M}$ from the bulk VB and high intensity bulk bands dispersing downwards toward $\Bar{M}$. \textbf{(c)} The bulk gap at the $\Bar{M}$ point and QW's of the bulk VB for \SI{10}{} and \SI{30}{BL} films at $k_x = \SI{0.78}{\AA^{-1}}$. The top of the VB can be seen to shift to lower energies in the \SI{10}{BL} film. \textbf{(d)} Narrow energy $\Bar{\Gamma}-\Bar{M}$ cut for for \SIrange{4}{30}{BL} Bi$_{0.85}$Sb$_{0.15}$ at $h\nu = \SI{37.5}{eV}$ and  $k_y = \SI{0.0}{\AA^{-1}}$. The reemergence of $S_1$ at the Fermi level happens between \SI{10}{} and \SI{30}{BL}. In the \SI{4}{} and \SI{10}{BL} films the surface states cross $E_F$ a total of 3 times and 5 times in \SI{30}{BL}, indicated by numbers. A decrease in the QW spacing with increasing thickness is shown. \textbf{(e)} MDC cuts taken at $E_F$ for the spectra shown in (d). There is a peak at \SI{0.6}{\AA^{-1}} for the \SI{30}{BL} film from the remergence of $S_1$ due to a reduced bulk gap.}
  \label{fig:EK}
\end{figure*}

Ultrathin  Bi$_x$Sb$_{1-x}$ films were grown using MBE on unintentionally doped InSb(111)B (Sb polar surface). Fig. \ref{fig:structure} shows that Bi$_{0.85}$Sb$_{0.15}$ films grow epitaxially, with a single orientation on InSb(111)B substrates, down to a few bilayers, enabling studies into the electronic structure in the two-dimensional limit.
The films studied were grown at a composition of Bi$_{0.85}$Sb$_{0.15}$. In Fig. \ref{fig:structure}(b), for three different thicknesses, XRD $2\theta-\omega$ scans of Bi$_{0.85}$Sb$_{0.15}$ show the observation of the Bi$_{0.85}$Sb$_{0.15}$(003) diffraction peak confirming the epitaxial nature of the film. As film thickness is increased from \SI{4}{} to \SI{100}{BL} the lattice fringe spacing decreases for the (003) peak, as expected for high quality, epitaxial films. Dynamical diffraction simulations reproduced the observed lattice fringes and film thickness \cite{Kriegner:rg5038}. The thickness of the films was also confirmed by RBS (supplemental information). A STM image of 4 BL Bi$_{0.85}$Sb$_{0.15}$, Fig. \ref{fig:structure}(c), shows terraces and steps of \SI{1}{} to \SI{2}{BL} with a single orientation (all triangles point in the same direction). Bi and Bi$_x$Sb$_{1-x}$ films grown on other substrates show rotational domains \cite{PhysRevMaterials.6.074204}. The in-plane lattice constant obtained from X-ray reciprocal space maps and confirmed with dynamical diffraction simulations gives an effective compressive strain of $2\%$ and the BiSb(0001)/[10$\Bar{1}$0] (hexagonal notation) $\parallel$ InSb(111)/[11$\Bar{2}$] epitaxial alignment, Fig. \ref{fig:structure}(a). However, in Van der Waals materials, such as Bi, the energy associated with elastic strain can be relieved by a rearrangement of atoms forming strain solitons \cite{Edelberg2020,Dong2024}. A random network of strain solitons are evident in Fig. \ref{fig:structure}(c) indicating the films are partially relaxed. The solitons are the metamorphic region and the strained part is pseudomorphic. 

\begin{figure*}
\centering
  \includegraphics[width=1\textwidth]{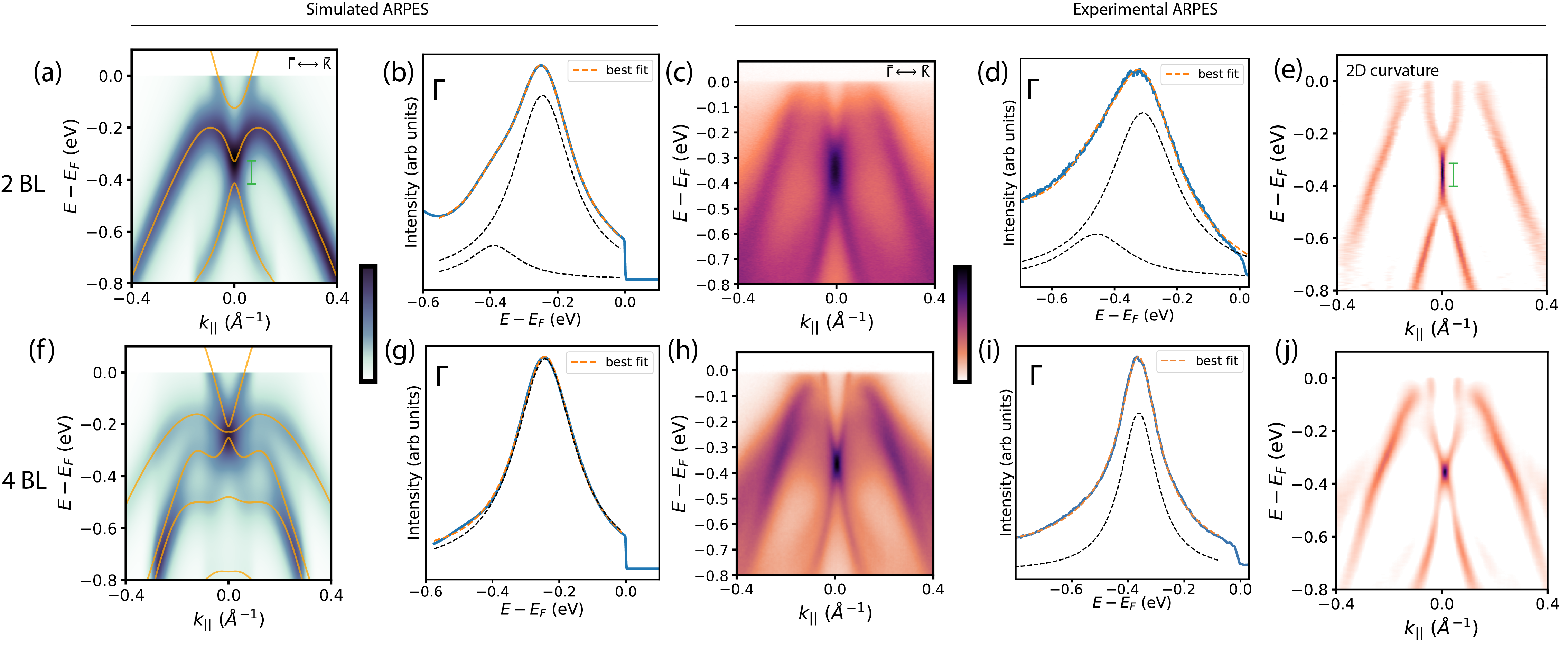}
  \caption{\textbf{Hybridization gap in ultrathin Bi$_{0.85}$Sb$_{0.15}$ films.} \textbf{(a)} Simulated ARPES intensity for a \SI{2}{BL} slab in the $\Bar{K}-\Bar{\Gamma}-\Bar{K}$ direction with $h\nu=\SI{37}{eV}$, $p$-polarized light, and a self-energy of \SI{50}{meV} at \SI{6}{K}. The bare dispersion is overlayed in orange. \textbf{(b)} A simulated EDC taken at the $\Bar{\Gamma}$ point. The best fit to the EDC is given by two Lorentzians indicating two seperate bands. The two Lorentzian components are represented by dotted black lines. \textbf{(c)} Measured ARPES spectra for a \SI{2}{BL} film in the $\Bar{K}-\Bar{\Gamma}-\Bar{K}$ direction with $h\nu=\SI{37}{eV}$, $p$-polarized light at \SI{6}{K}. The spectral intensity is broad in energy at $\Bar{\Gamma}$. \textbf{(d)} An experimental EDC taken at the $\Bar{\Gamma}$ point. There is peak and a shoulder deeper in binding energy. The best fit to the simulated EDC is given by two Lorentzians indicating two separate bands as evidence for a hybridization gap in the TSS. \textbf{(e)} 2D curvature plot emphasizing band locations. Intensity in the gap at $\Gamma$ is due to either residual MDC intensity peak at $\Gamma$ or broad EDC peaks. \textbf{(f)} Simulated ARPES intensity for a \SI{4}{BL} slab in the $\Bar{K}-\Bar{\Gamma}-\Bar{K}$ direction. \textbf{(g)} A simulated EDC taken at the $\Bar{\Gamma}$ point. There is sharp peak and the best fit to the simulated EDC is given by a single Lorentzian where the bands get close to touching. \textbf{(h)} Measured ARPES spectra for a \SI{4}{BL} film in the $\Bar{K}-\Bar{\Gamma}-\Bar{K}$ direction with $h\nu=\SI{37}{eV}$, $p$-polarized light at \SI{6}{K}. The spectral intensity is sharp in energy at $\Bar{\Gamma}$. \textbf{(i)} An experimental EDC taken at the $\Bar{\Gamma}$ point. There is single, sharp peak in binding energy. The best fit to the measured EDC is given by a single Lorentzian indicating no resolved gap within the energy resolution of the experiment. \textbf{(j)} 2D curvature plot emphasizing band locations. The gap at $\Gamma$ is not detected with the resolution of the experiment.}
  \label{fig:hybrid}
\end{figure*}

The high quality growth of ultrathin Bi$_x$Sb$_{1-x}$ on InSb(111)B makes it a prime material to study the effects of confinement on the electronic structure. 
A tight-binding model incorporating matrix elements was used to simulate ARPES measurements at various thicknesses \cite{PhysRevB.52.1566,Day2019} (see supplemental for details). 
Since the TSS in Bi$_x$Sb$_{1-x}$ connect the bulk VB and CB at the $L$ point, reduced dimensionality should lead to an increased bulk gap and a larger separation in energy of the TSS. Fig. \ref{fig:structure}(d) shows the quantization of the bulk VB into QW with higher confinement energy deeper in binding energy. The top of the VB moves to deeper binding energy at the $\Bar{M}$ point with decreasing thickness, and the TSS separate in energy on the order of \SI{}{eV} since they connect to the VB and CB. Scattering with unpolarized bulk states can depolarize the TSS therefore larger energy separation from the bulk states is sought after\cite{PhysRevLett.106.257004,PhysRevLett.123.207205}. Fig. \ref{fig:structure}(e) shows the spectral function for two slabs \SI{4}{} and \SI{10}{BL} thick. 
In the \SI{4}{BL} film both branches of the spin-polarized topological surface state ($S_1$ inner and $S_2$ outer at $+k$, $S_1$ outer and $S_2$ inner at $-k$) are visible with $S_1$ connecting to the CB and $S_2$ connecting to the VB at the bulk $L$ point. As the thickness of the slab is increased, the bulk gap is reduced and so is the separation of the TSS from eachother, seen in the \SI{10}{BL} film. In thick films as the bulk gap gets smaller, there is a reemergence of $S_1$ at the Fermi level (supplemental materials).
The number of QW states originating from the bulk can be seen to increase from the \SI{4}{} to \SI{10}{BL} systems. 

\begin{figure*}
\centering
  \includegraphics[width=1\textwidth]{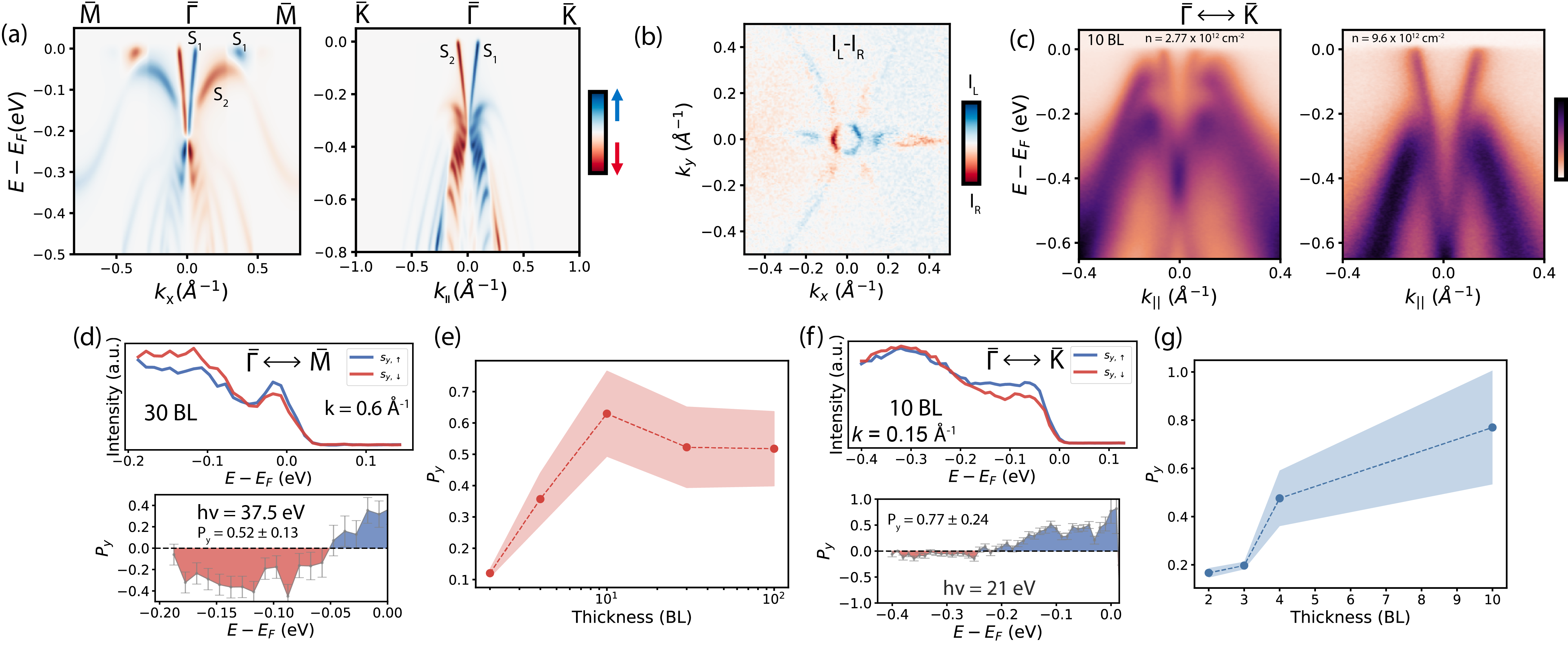}
  \caption{\textbf{Spin-polarization of TSS in ultrathin Bi$_{0.85}$Sb$_{0.15}$.} \textbf{(a)} Simulated spin-resolved ARPES intensity for a \SI{15}{BL} slab in the $\Bar{M}-\Bar{\Gamma}-\Bar{M}$ and $\Bar{K}-\Bar{\Gamma}-\Bar{K}$ direction with $h\nu=\SI{21}{eV}$, $p$-polarized light at \SI{14}{K}. The spin is projected onto the $y$-axis with $S_1$ having opposite spin to $S_2$. \textbf{(b)} Experimental difference in photoemission intensity using circularly polarized light at $E_F$. The inner branch of the TSS shows an opposite circular dichroism signal across $\Bar{\Gamma}$. \textbf{(c)} A \SI{10}{BL} film of Bi$_{0.85}$Sb$_{0.15}$ taken at $h\nu = \SI{21}{eV}$ before and after electron doping from K-dosing. \textbf{(d)} Spin-resolved EDC and $P_y$ for a \SI{30}{BL} film in the $\Bar{\Gamma}-\Bar{M}$ direction at \SI{0.6}{\AA^{-1}} and $h\nu=\SI{37}{eV}$. Spin is projected onto the $y$-axis. The spin-polarization of $S_1$ at the Fermi level is evident in the \SI{30}{BL} film with a $P_y$ of $0.5$ for the lower branch of the TSS, $S_2$. \textbf{(e)} $P_y$ for all films measured in the $\Bar{\Gamma}-\Bar{M}$ direction at \SI{0.6}{\AA^{-1}}. Error bars indicated by shaded region. \textbf{(f)} Spin-resolved EDC and $P_y$ for a \SI{10}{BL} film in the $\Bar{\Gamma}-\Bar{K}$ direction at \SI{0.15}{\AA^{-1}} and $h\nu=\SI{21}{eV}$. Spin is projected onto the $y$-axis. There is spin-polarization from the TSS before it merges into the bulk.  \textbf{(g)} $P_y$ for all films measured in the $\Bar{\Gamma}-\Bar{K}$ direction at \SI{0.15}{\AA^{-1}}. The \SI{10}{BL} film shows a very large $P_y$ of $0.77 \pm 0.24$.}
  \label{fig:spin}
\end{figure*}

Fig. \ref{fig:EK}(a) shows the Fermi surface of a \SI{30}{BL} film taken at $h\nu=$\SI{37.5}{eV}, corresponding to a $k_z$ of the bulk $\Gamma$ point \cite{PhysRevB.70.245122}. The Fermi surface at \SI{37.5}{eV} is three-fold symmetric corresponding to the $C_{3v}$ symmetry of the crystal structure, consistent with single orientation films. There are lower intensity states rotated by \SI{60}{\degree} could be due to the fact that the surface is six-fold symmetric. At the Fermi level there are both TSS present, $S_1$ and $S_2$, in the $\Bar{\Gamma}-\Bar{M}$ direction. 
Taking a $\Bar{\Gamma}-\Bar{M}$ cut in a \SI{30}{BL} film at \SI{81}{eV}, Fig. \ref{fig:EK}(b), shows both TSS disperse through the Fermi level near $\Bar{\Gamma}$ then remerge as $S_2$ connects to the VB at the bulk $L$ point and $S_1$ connects to the CB. 
This dispersion is generally consistent with previous reports for bulk BiSb alloys \cite{doi:10.1126/science.1167733}.
Discussion regarding which surface state connects to the CB and related mirror Chern number can be found in Ref. \cite{PhysRevB.78.045426}. In addition to the surface states we see high intensity bulk states around \SI{-1}{eV} at $\Bar{\Gamma}$ dispersing downwards towards $\Bar{M}$. 
As the thickness is reduced the bulk gap size increases, Fig. \ref{fig:EK}(c) shows the dispersion of the VB at $\Bar{M}$ in the $k_y$ direction and EDC's for \SI{10}{} and \SI{30}{BL}. The VB maximum sits right below $E_F$ in the \SI{30}{BL} film but shifts to $\sim \SI{160}{meV}$ below $E_F$ in the \SI{10}{BL} film, as well as the QW spacing of the VB increases with reduced thickness. Fig. \ref{fig:EK}(c) provides direct evidence the bulk gap increases since the first peak in the \SI{10}{BL} EDC is lower in deeper in binding energy.
In Fig. \ref{fig:EK}(d) ARPES spectra in the $\Bar{\Gamma}-\Bar{M}$ direction for \SI{4}{BL}, \SI{10}{BL}, and \SI{30}{BL} are shown highlighting the progression of the electronic structure with decreasing confinement effects. As thickness is increased there is a decrease in the spacing of the QW states and a reemergence of $S_1$ at the Fermi level. Alternatively, momentum distribution curves (MDC) at $E_F$ show the reemergence of $S_1$ at \SI{0.6}{\AA^{-1}} resulting from the reduced bulk gap. In all three thickness we see the surface states cross the Fermi level an odd number of times (3 in \SI{4}{},\SI{10}{BL} and 5 in \SI{30}{BL}) confirming the nontrivial topology of the bulk. Our ARPES results indicate that confinement of our high quality films can increase the bulk gap and separation of the TSS.

When the thickness of the films are reduced to a few bilayers, the TSS wave functions will begin to overlap and hybridize, leading to a gap in the Dirac point since the bands of the same spin will avoid a crossing \cite{Zhang2010}. It has been proposed that a three dimensional topological insulator can transition into a two dimensional quantum spin Hall state when thickness is reduced to the point of opening a hybridization gap \cite{PhysRevB.81.041307}. Therefore, utilizing high quality MBE grown films to observe evidence of a gap in the TSS in ultrathin Bi$_{0.85}$Sb$_{0.15}$ would open up the material system for studying the QSHE. By projecting the eigenstates of the Hamiltonian onto the orbitals at the surface of the slab, the penetration depth of the TSS can be estimated. The TSS have significant orbital weight in the first bilayer, with some in the second (see supplemental information). Therefore we would expect to see evidence for a hybridization gap in the ARPES spectra of a \SI{2}{BL} film. Fig. \ref{fig:hybrid}(a) shows a simulated ARPES spectrum for a \SI{2}{BL} film in the $\Bar{\Gamma}-\Bar{K}$ direction, accurately reproducing the experimental spectrum shown in Fig. \ref{fig:hybrid}(c) with the bare bands overlayed clearly showing a hybridization gap. The ARPES spectrum is simulated at \SI{6}{K} with a self energy of \SI{50}{meV} resulting in a spectrum that has no obvious gap. By taking a EDC at the $\Bar{\Gamma}$ point, Fig. \ref{fig:hybrid}(b), it is clear two bands are present by the long shoulder plus small hump at \SI{400}{meV} below $E_F$. This is confirmed by the best fit of the linecut with two Lorentzians, each centered at the respective band locations. The measured spectrum (Fig. \ref{fig:hybrid}(c)) and corresponding EDC at $\Bar{\Gamma}$ (Fig. \ref{fig:hybrid}(d)) display similar behavior. The spectral intensity at $\Bar{\Gamma}$ is broader in energy which is clear in the linecut which has a broad shoulder lower in binding energy. The EDC is best fit with two Lorentzians rather than one (lower $\chi ^2$) separated in energy by the same amount as the simulated ARPES spectra, presenting evidence for two bands and a hybridization gap. This observation is in contrast to a \SI{4}{BL} film where the gap in the surface states is negligibly small leading to sharp spectral intensity at $\Bar{\Gamma}$, Fig. \ref{fig:hybrid}(f,g). The EDC's are both best fit by a single Lorentzian, Fig. \ref{fig:hybrid}(h,i). Band locations and gaps are made more clear in the 2D curvature plots Fig. \ref{fig:hybrid}(e,j), emphasising the opening of a gap in the few bilayer limit \cite{10.1063/1.3585113}. Our measured ARPES spectrum for an ultrathin \SI{2}{BL} film of Bi$_{0.85}$Sb$_{0.15}$ is evidence for a hybridization gap, suggesting MBE growth of BiSb alloys is a viable route to study the QSHE. 

Understanding how the helical spin texture of the TSS changes with thickness provides a means to probe hybridization, and is important for devices especially in the ultrathin limit. 
Fig. \ref{fig:spin}(a,b) shows the simulated spin-resolved ARPES spectral intensity of a \SI{15}{BL} slab. 
In the $\Bar{\Gamma}-\Bar{M}$ direction, we see both $S_1$ pockets are of opposite spin to $S_2$ when projected onto the $y$-axis. 
In the $\Bar{\Gamma}-\Bar{K}$ direction the TSS that linearly disperse and connect to the CB are of opposite spin. 
Fig. \ref{fig:spin}(b) displays the experimental Fermi surface captured using circularly polarized light integrated from the Fermi level to \SI{5}{meV} below the Fermi level. The inner branch of the TSS ($S_1(+k)$+$S_2(-k)$) exhibits momentum-dependent circular dichroic (CD) photoemission intensity with the sign switched across the $\Bar{\Gamma}$ point, inline with what is expected for the Dirac cone of a topological insulator \cite{https://doi.org/10.1002/pssr.201206458}. The $S_2$ pocket switches sign across the mirror plane from $\Bar{\Gamma}-\Bar{M}$. The CD signal depends on experimental geometry, with photon and emission plane orientation relative to the mirror plane determining if the signal is odd or even in momentum, therefore we use SARPES to measure the magnitude of the spin polarization.
Accessing the TSS above the Fermi level in the $\Bar{\Gamma}-\Bar{K}$ direction allows us to better study the spin-polarization within the energy and momentum resolution of the analyser and spin detector. Fig. \ref{fig:spin}(c) shows a $\Bar{\Gamma}-\Bar{K}$ cut for a \SI{10}{BL} film both pre and post Potassium dosing which electron dopes the sample \cite{PhysRevB.97.155423}. The inner branch of the TSS are both clearly seen linearly dispersing upwards. 
A Sherman function (S) of $0.24$ was used. The spin-polarization was taken along the $y$-axis, $P_y = (1/S) \ast (I_{\uparrow}-I_{\uparrow})/(I_{\uparrow}+I_{\uparrow})$, to measure the helical nature of the TSS. A spin-resolved EDC for \SI{30}{BL} before K-dosing at \SI{0.6}{\AA^{-1}} in the $\Bar{\Gamma}-\Bar{M}$ direction with corresponding $P_y$ is shown in Fig. \ref{fig:spin}(d). In the thicker film there is a significant $P_y$ at the Fermi level from where $S_1$ crosses the Fermi level for the second and third time. $P_y$ then switches sign as a function of binding energy as the spin-polarization comes from $S_2$. 
$P_y$ for all thicknesses cut at \SI{0.6}{\AA^{-1}} in the $\Bar{\Gamma}-\Bar{M}$ direction is shown in Fig. \ref{fig:spin}(e). $P_y$ peaks at \SI{10}{BL} and then drops as the thickness is further reduced. The peak at \SI{10}{BL} is likely because quantum confinement has increased the bulk gap, therefore the TSS are separated in energy from the bulk states leading to less spin scattering. Below \SI{10}{BL} hybridization effects begin to reduce the spin-polarization. 
Fig. \ref{fig:spin}(f) shows a spin EDC and $P_y$ for \SI{10}{BL} taken in the $\Bar{\Gamma}-\Bar{K}$ direction after K-dosing. As a function of binding energy, $P_y$ decreases from the Fermi level as the TSS begin to connect to the VB. At the film thickness of \SI{10}{BL} $P_y$ has a very large value of $0.77 \pm 0.24$. To the best of our knowledge, this is the highest reported value for any Bi$_{1-x}$Sb$_{x}$ or Bi system \cite{doi:10.1126/science.1167733,Takayama2012}. Additionally, the spin-polarization observed in the \SI{2}{BL} film is the thinnest topological insulator sample reported to have any measurable spin polarization in SARPES. We see that by reducing film thickness quantum confinement effects increase the bulk band gap which increases the energy seperation of the topological surface states leading to an approaching unity spin-polarization in ultrathin Bi$_{0.85}$Sb$_{0.15}$. 
The reduction in spin-polarization with thickness, Fig. \ref{fig:spin}(g), signals hybridization of the TSS in ultrathin films.

\section*{Conclusion}

In summary, we have epitaxially grown the thinnest Bi$_{1-x}$Sb$_{x}$ films made possible by the appropriate choice of InSb(111)B as the substrate. The films were epitaxial down to \SI{2}{BL} with a single orientation. We have characterized the electronic structure showing that confinement in thin films increase the bulk band gap and changes the number of times the topological surface states cross the Fermi level. In ultrathin films we observe a gap in the topological surface states from hybridization, opening the door for investigations into the QSHE. The topological surface states show a spin-polarization approaching unity as the bulk band gap increases before the surface states hybridize, making Bi$_{1-x}$Sb$_{x}$ films suitable for spintronic applications. Our films provides a means to study well known layered topological systems in a new regime, opening up previously understood materials for use in scalable topological devices, by approaching the two-dimensional limit with MBE.

\section*{Acknowledgements}

The growth and later ARPES studies were supported by the U.S. Department of Energy (contract no. DE-SC0014388). The initial vacuum suitcase construction and initial ARPES measurements were supported by the US Department of Energy (DE-SC0014388). The UC Santa Barbara NSF Quantum Foundry funded via the Q-AMASE-i program under award DMR-1906325 support was used for further development of the vacuum suitcases. This research used resources of the Advanced Light Source, which is a DOE Office of Science User Facility under contract no. DEAC02-05CH11231. The research reported here made use of the shared facilities of the Materials Research Science and Engineering Center(MRSEC) at UC Santa Barbara: NSFDMR–2308708.

\section*{Author contributions}
P.C., C.J.P conceived the project. P.C. grew the films. P.C. and A.E. performed the ARPES/SARPES. P.C. performed the XRD. P.C. and J.D. performed the STM. P.C. wrote the tight-binding model. P.C. performed the data analysis. The manuscript was written by P.C. and C.J.P. All authors contributed in discussing the results and forming conclusions.

\bibliographystyle{apsrev4-2}
\bibliography{bib.bib}

\end{document}